# Deep learning-based citation recommendation system for patents

Jaewoong Choi[a], Sion Jang[b], Jaeyoung Kim[b], Jiho Lee[a], Janghyeok Yoon[a,*], Sungchul Choi[b,*]

[a]Department of Industrial Engineering, Konkuk University, Seoul 05029, Republic of Korea

[b]TEAMLAB, Department of Industrial and Management, Gachon University, Seongnam-si, Gyeonggi-do, Republic of Korea

*Corresponding author: janghyoon@konkuk.ac.kr, sc82.choi@gachon.ac.kr

Abstract

In this study, we address the challenges in developing a deep learning-based automatic patent citation recommendation system. Although deep learning-based recommendation systems have exhibited outstanding performance in various domains (such as movies, products, and paper citations), their validity in patent citations has not been investigated, owing to the lack of a freely available high-quality dataset and relevant benchmark model. To solve these problems, we present a novel dataset called PatentNet that includes textual information and metadata for approximately 110,000 patents from the Google Big Query service. Further, we propose strong benchmark models considering the similarity of textual information and metadata (such as cooperative patent classification code). Compared with existing recommendation methods, the proposed benchmark method achieved a mean reciprocal rank of 0.2377 on the test set, whereas the existing state-of-the-art recommendation method achieved 0.2073.

## 1. Introduction

As technology advances rapidly, an exponentially growing number of patents confront patent examiners and applicants with information overload problems. Patent applicants may retrieve numerous irrelevant patents when trying to find citable patents. In case of the experts, searching related patents is crucial in designating appropriate citable patents during patent examination. Therefore, patent citation recommendation systems are required to provide an effective solution to the aforementioned practical problem. Although citation recommendation systems have demonstrated successful recommendation examples [1], their application to patents may not yield satisfying results because of not considering the properties of patent citation contexts.

Several pioneering researchers have proposed automatic patent citation recommendation systems, leaving room for performance improvement [2, 3]. First, the pioneers did not use a patent citation dataset labeled by examiners (i.e., patent experts), because patent citation recommendation is a careful task that may affect the legal scope of a citing patent and its further economic implications. In addition, patent examiners responsible for adjudicating the patentability of a citing patent examine whether the applicant citation is correct and determine the cited patents appropriately. Therefore, the examiner citations dataset is more accurate and suitable than the applicant citations in prior studies. Second, a patent citations decision considers the content and the technical, industrial, and legal relevance between the citing and cited patents [4]. For example, metadata (e.g., technology classification codes) is an invaluable feature to understand the patent citation context. However, prior studies focused only on textual information for recommendation. Finally, it is difficult to compare other existing methods because of the absence of a freely available dataset, making the experimental results less valuable.

In this study, we addressed the challenges in developing an automatic patent citation recommendation system. We proposed a novel dataset, called PatentNet, labeled by patent examiners, and it includes textual information and metadata for approximately 110,000 patents collected from the Google Big Query service. Further, we proposed a strong deep learning-based benchmark model to consider the similarity of textual information and metadata. The proposed model consists of two different stages. In the first stage, the patent features were fed into a dense embedding by encoding the textual information and pre-trained metadata information. In the next stage, the top-K nearest neighbor patents of a query patent were selected and re-ranked using a re-ranking model. The main contributions of this study can be described as follows.

- We propose a deep learning-based patent citation recommendation system, in which patent textual information and pre-trained metadata are used to represent individual patents. To the best of our knowledge, this is the first study in which both textual information and metadata were employed in patent citation recommendation systems.
- We propose a novel dataset called PatentNet as a patent citation recommendation dataset, which includes textual information (title, abstract) and metadata for approximately 110,000 patents. We expect researchers to conduct further studies regarding patent citation with this dataset.
- We design a patent citation recommendation model that is applicable to new incoming patents. Our model can reproduce citation recommendations for new patents that are not used in both training and testing stages.

The remainder of this paper is organized as follows. Section 2 presents related work. Section 3 describes the proposed dataset and the structure of our model. The experimental and evaluation results are presented in Section 4. Section 5 concludes the paper and provides potential future research directions.

2. Related works
2.1 Citation recommendation

Citation recommendations help users to find citable items for their personalized information. Several models have been proposed to implement a citation recommendation task. In general, citation recommendation systems are divided into two different types based on their usage pattern: global citation recommendation and local citation recommendation. First, global citation recommendation considers the entire manuscript as a query paper to generate a list of citation recommendation. Since Gipp, Beel and Hentschel [5] developed the initial global citation recommendation system, various models have been proposed. Mu, Guo, Cai and Hao [6] proposed a multi-layered, mutually reinforced query-focused citation recommendation approach, where multiple types of relations between authors, papers, and keywords were used to represent the global contextual information. Tian and Zhuo [7] utilized distributed word representations learned from the global citation context to measure the similarities of papers. An attention-based convolutional neural network was used by Du, Tang and Ding [8] for calculating the relevance between a query paper and candidates to solve a personalized recommendation problem. Ma and Wang [9] jointly used author and paper profiles to represent heterogenous graphs, thereby solving the problem of the personalized recommendation problem. Yang, Zhang, Cai and Guo [10] proposed an efficient recommendation model by considering the problem as a hidden edge prediction in a heterogeneous bibliographic network, which was based on paper, author, and venue information. Cai, Zheng, Yang, Dai and Guo [11] combined bibliographic network information with author, paper, and venue contents, thereby enabling personalized citation recommendation for a given query paper.

The second type, local citation recommendation (also known as context-aware citation recommendation), considers only the citation context, which is the text where the citation is located, as a query. That is, these models focus only on sentences surrounding a placeholder that represents a citation. He, Pei, Kifer, Mitra and Giles [12] first introduced a context-aware citation recommendation system, where a non-parametric probabilistic model was employed to recommend citations for a context. Ebesu and Fang [13] attempted to embody a robust representation of citation context by utilizing author networks and semantic composition of citation contexts. Yang, Zheng, Cai, Dai, Mu, Guo and Dai [14] incorporated author information, venue information, and contents to represent papers, thereby increasing the recommendation performance. Jeong, Jang, Park and Choi [1] employed graph convolutional network layers and bidirectional encoder representations from transformers to represent the citation graph and context, respectively. Zhang and Ma [15] attempted to embed academic papers as dense vectors using contents, local context, and structural context. Although there exists no absolutely correct method for citation recommendation, the global citation recommendation is being adopted more than the local one [16]. In this study, our main focus is the global citation recommendation problem.

The factors used in our explored models can be summarized into contents, tagged keywords, author information, venue information, citation network, and social network. The factors are employed in the form of a matrix or graph to represent documents. In particular, the best part of the models exploited the paper contents involving the abstract, title, and manuscript as the main source to represent a query paper. In addition, tagged keywords are frequently used to identify user interests. User profile and venue information are sometimes employed to boost recommendation performance. Citation network or social network have also been frequently used to reflect the relations between papers and authors, which are expected to refine the recommendation results.

2.2 Patent features for citation recommendation

Patent citation recommendation models can benefit from various patent characteristics. Because patents are intellectual property managed by the official patent office, the patent data is stored in a structured form [17]. Accordingly, patent data include structured contents such as bibliographic information, metadata, and citations. The technical scope can be determined by the title and abstract of a patent, which describe the technical contents in detail [18]. In addition, all patents are assigned with multiple technology classification codes, such as international patent classification (IPC) code and cooperative patent classification (CPC) code [19]. Such codes indicate the industry or technologies involved in each patent, and thus, they have been used to calculate the similarities between patents or applicants [20]. In this study, textual information obtained from the title and abstract, as well as the CPC code information, are employed to construct a patent citation recommendation model.

3. Proposed dataset and model
3.1 Proposed dataset

For patent citation recommendation tasks, more generalized and robust Deep Neural Networks (DNNs) can be proposed by exploiting a high-quality patent dataset, resulting in better applications for users to recommendation with this dataset. We believe such dataset is a critical resource to develop advanced, content-based patent recommendation and classification algorithms, as well as to provide critical training and benchmarking data for such algorithms.

As shown in Fig. 1, previous studies used uncertain datasets that were not confirmed by examiners; by contrast, PatentNet is built upon accurate patents verified by examiners. To obtain the examiner dataset, we first collected patents from the KISTA website, which contained patent technology reports determined by patent examiners with patent search expressions. Subsequently, we transformed the search expression to fit the Google Big Query database and extracted a total of 800,000 geostationary orbit complex satellite (GOCS) field patents. Finally, we selected approximately 110,000 patents with mutual citation relations and intact features (such as title, abstract, CPC, IPC, USPC, and registration year); that is, these patents were reprocessed to create a dataset for patent citation recommendation.

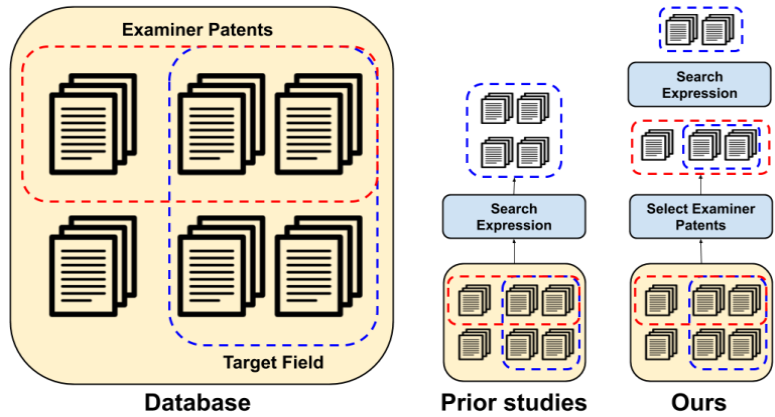

**Figure 1.** Overall procedure for PatentNet

The statistics of the proposed dataset are presented in Table 1. Base patents refer to individual patents that cite other patents. In the proposed dataset, a total of 43,073 base patents cites totally 67,252 patents and generates an overall of 159,157 citation relations. The registration year interval of the collected patents ranges from 1971 to 2016. The proposed dataset includes textual information, such as title and abstract for the total patents. In addition, the dataset contains metadata such as CPC, IPC, and USPC codes. The number of classification codes is obtained at the subclass level. Based on the proposed dataset, we defined the patent citation relationships that were determined by patent examiners.

**Table 1.** Data statistics

| Feature | Value |
| --- | --- |
| # of total patents | 107,087 |
| # of base patents | 43,073 |
| # of cited patents | 67,252 |
| # of citation number | 159,157 |
| Patent registration year | 1971-2016 |
| # of CPC code | 152,909 |
| # of IPC code | 50,027 |
| # of USPC code | 104,835 |

3.2 Proposed model

As motivated by Bhagavatula, Feldman, Power and Ammar [21] the overall framework of the proposed method consists of two stages. During the first stage, we trained a candidate selection network (CSNet) to extract the candidate patents corresponding to the query patent. In the next stage, a candidate reranking network (CRNet) was trained to rank those citationable patents. We used the CPC embedding layer in our CRNet to achieve further improved performance. Detailed descriptions regarding these procedures are presented in the following sections. Meanwhile, the experimental results section also provides analysis regarding the contribution of each stage to the final patent recommendation performance.

3.2.1 Candidate Selection Network

The proposed framework aims to provide citationable patents; however, its performance is insufficient to analyze all dataset patents against query patents during inference. We trained the CSNet to identify a candidate set for citationable patents for queries without explicitly iterating over all documents in the database. The CSNet tries to project any patent $p$ to a dense embedding based on its contents. To

construct the patent feature vectors, we aggregate the title, abstract, and CPC code:

$$e = \lambda_t \sum_{i=1}^{S_t} g(p_{title}) + \lambda_a \sum_{i=1}^{S_a} g(p_{abstract}) + \lambda_c g(p_{cpc}), \ p \in \mathbb{R}^E \quad (1)$$

where $S_t$ and $S_a$ denote the sequence length of the title and abstract, respectively, while $g(\cdot)$ indicates weight normalization, and $\lambda$ is a randomly initialized trainable scalar parameter. We use a bag-of-words representation of each field. $E$ is the dimension of feature vectors, which is set to 75. To train $f(p)$, the triplet loss with $l2$ regularization is used for each field, which is computed as:

$$loss = \max(1 + sim(e_q, e^-) - sim(e_q, e^+), 0) + \beta \sum_k w_k^2 \quad (2)$$

where, $e_q$ refers to a query patent feature, while $e^+$ and $p^-$ denote the patent cited by $e_q$ and that is not, respectively. $sim(\cdot)$ is a cosine-similarity, $w$ is a set of trainable weights, and $\beta$ is a regularization parameter, which was set to $10^{-4}$. The CSNet was trained for a total of 50 epochs with the Adam optimizer and an initial learning rate of $10^{-4}$. During the inference, we selected $K$-nearest neighbor patents in the embedding space as our candidates.

3.2.2 Candidate Reranking Network

In the patent citation recommendation, reflections of the patent unique features are important as they may affect the citation relationship. We employed inputs to facilitate the relationship inference between query patents and candidates, as illustrated in Fig. 2.

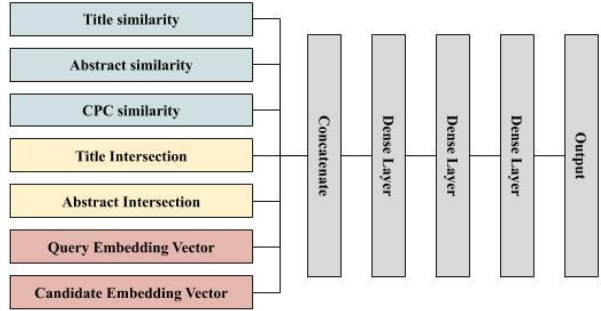

**Figure 2.** Architecture of Candidate Reranking Network

For the input representation, we used numeric and word intersection features, as well as embedding features, which can be expressed as follows:

$$x = x_{num} \oplus x_{int} \oplus x_{emb} \quad (3)$$

$$x_{num} = sim(p_{title}^q, p_{title}^c) \oplus sim(p_{abstract}^q, p_{abstract}^c) \oplus sim(p_{cpc}^q, p_{cpc}^c) \quad (4)$$

$$x_{int} = \sum_{i \in Title} w_i \oplus \sum_{i \in Abstract} w_i \quad (5)$$

$$x_{emb} = e^q \oplus e^c \quad (6)$$

where $\oplus$ is the concatenate operator, $x_{num}$ is the numeric feature computed as using the cosine similarity in query and candidate for title, abstract, and CPC, respectively. $x_{int}$ represents the weight summations of the intersection words in the query $p^q$ and candidate $p^c$. $x_{emb}$ is the concatenate query embedding vectors $e^q$ and candidate $e^c$. The CRNet $f(\cdot)$ consists of totally 4 dense layers. The last layer of CRNet used sigmoid activation function and the other layers used elu. To train the CRNet, the binary cross entropy loss is used:

$$loss = \frac{1}{K}\sum_{k=1}^{K} y^t \log\big(f(p^q, p_k^c)\big) + (1 - y^t)\log\big(1 - f(p^q, p_k^c)\big) \tag{7}$$

where $y^k$ represents the ground truth label, which is annotated as 1 if the candidate patent is cited for the query. $K$ indicates the number of candidate patents with high similarity to the query patent in the CSNet embedding space. The nodes of each layer for CRNet consists of 20, 20, 20, and 1, respectively. The CRNet was trained for a total of 400 epochs with the Adam optimizer and an initial learning rate of $10^{-2}$ obtained from hyperparameter tuning.

4. Experiments and evaluations

4.1 Experiment overview

We conducted experiments on the PatentNet. Moreover, we also compared the proposed model with BM25 and carried out an ablation study to quantitatively assess the contribution of each step in our proposed training framework. We employed Recall@K, Precision@K, F1-score@K, and the mean reciprocal rank (MRR) as our evaluation metrics. The experiment targets to achieve the following detailed goals, as well as overall performance improvement of patent citation recommendations.

- Show the superior performance of our model that reflects the patents' unique characteristics, through comparison with the existing recommendation models, such as BM25.
- Investigate the potential effect of using pre-trained meta information such as CPC, which is a unique characteristic of patents, as an additional feature.

4.2 Experiment setting

4.2.1 Evaluation metrics

In this section, we used MRR, Recall@K, Precision@K and F1@K to evaluate the experimental results, which are the most popular evaluation metrics used in citation recommendation systems. F1@K is calculated as the harmonized average of Recall@K and Precision@K. Here, @K refers to the values in a top $K$ recommendation list. Accordingly, F1@K is defined as:

$$F1@K = \frac{2 * Recall@K \times Precision@K}{Recall@K + Precision@K} \tag{8}$$

The MRR indicator computes the ranking reciprocal of the first relevant document retrieved per query, then averages them across all queries. That is, MRR evaluates the models that return a ranked list of citable patents corresponding to a set of query patents.

4.2.2 Experimental Results

As shown in Table 2, the compared method was 0.0808 in Recall, 0.0508 in Precision, 0.0624 in F1-score, and 0.1641 in MRR. The CSNet and CRNet, trained by using title and abstract features showed comparable performance comparing to the BM25 approach. Moreover, to evaluate the corresponding contribution, we experimented with different subsets of the CPC features used in CSNet and CRNet. By adding CPC features, Recall@20 of CRNet increased from 0.221 to 0.2232, while that of CSNet increased from 0.1572 to 0.1754. Also, we observed that when the proposed method included CPC features, CRNet outperformed the compared methods regarding all evaluation metrics. The CRNet with CPC features showed the superior performance (Recall@20: 0.2232, Precision@20: 0.1429, F1-Score@20: 0.1742, MRR: 0.234).

**Table 2.** Performance comparison of the proposed method on the test dataset for PatentNet

| Method | Recall@20 | Precision@20 | F1-Score@20 | MRR |
|---|---|---|---|---|
| BM 25 | 0.0808 | 0.0508 | 0.0624 | 0.1641 |
| CSNet w/o CPC | 0.1572 | 0.1013 | 0.1232 | 0.21 |
| CRNet w/o CPC | 0.221 | 0.1406 | 0.1719 | 0.2262 |
| CSNet with CPC | 0.1754 | 0.1129 | 0.1374 | 0.2268 |
| CRNet with CPC | **0.2232** | **0.1429** | **0.1742** | **0.234** |

We reported the Recall@K to investigate the performance change according to $K$, which are illustrated in Table 3. By adjusting the value of $K$ from 10 to 50, we compared the performance of our model that reflects the patent's unique characteristics such as CPC features with that of the existing recommendation model of BM25. As a result, regarding all observed $K$ values, our model showed the superior performance in terms of Recall compared to the existing model. Also, by adjusting the value of $K$, we investigated the potential effect of the pre-trained information (i.e., CPC features). We observed that the model with CPC features outperformed the models without CPC features regarding all observed $K$ values. It is obviously our model performs better than the compared model on all evaluation metrics and the detailed performance of Precision@K and F1-score@K can be found in Table 3-5.

**Table 3.** Recall@K comparison of the proposed method on the test dataset for PatentNet

| Method | Recall@10 | Recall@20 | Recall@30 | Recall@40 | Recall@50 |
|---|---|---|---|---|---|
| BM25 | 0.0555 | 0.0508 | 0.0958 | 0.104 | 0.1119 |
| CSNet w/o CPC | 0.1027 | 0.1572 | 0.1952 | 0.2202 | 0.2397 |
| CRNet w/o CPC | 0.1183 | 0.221 | 0.2888 | 0.3306 | 0.3666 |
| CSNet with CPC | 0.1122 | 0.1754 | 0.214 | 0.2348 | 0.2522 |
| CRNet with CPC | **0.1285** | **0.2232** | **0.2979** | **0.3454** | **0.3789** |

**Table 4.** Precision@K comparison of the proposed method on the test dataset for PatentNet

| Method | Precision@10 | Precision@20 | Precision@30 | Precision@40 | Precision@50 |
|---|---|---|---|---|---|
| BM25 | 0.0691 | 0.0508 | 0.0403 | 0.0329 | 0.0283 |
| CSNet w/o CPC | 0.1317 | 0.1013 | 0.084 | 0.0714 | 0.0624 |
| CRNet w/o CPC | 0.148 | 0.1406 | 0.1234 | 0.1064 | 0.0947 |
| CSNet with CPC | 0.1435 | 0.1129 | 0.0921 | 0.076 | 0.0652 |
| CRNet with CPC | **0.1614** | **0.1429** | **0.128** | **0.1115** | **0.0979** |

**Table 5.** F1-score@K comparison of the proposed method on the test dataset for PatentNet

| Method | F1-score @10 | F1-score @20 | F1-score @30 | F1-score @40 | F1-score @50 |
|---|---|---|---|---|---|
| BM25 | 0.0616 | 0.0624 | 0.0567 | 0.05 | 0.0452 |
| CSNet w/o CPC | 0.1154 | 0.1232 | 0.1175 | 0.1078 | 0.099 |
| CRNet w/o CPC | 0.1315 | 0.1719 | 0.1729 | 0.1609 | 0.1506 |
| CSNet with CPC | 0.1259 | 0.1374 | 0.1289 | 0.1148 | 0.1037 |
| CRNet with CPC | 0.1431 | 0.1742 | 0.179 | 0.1686 | 0.1557 |

We also presented some examples of qualitative recommendation results of a patent (US8612668B2), to deeply understand how our model performs better than the compared model. As shown in Table 6, the title and abstract of the patent addressed an invented technology regarding storage optimization system based on object size and its CPC was G06F (ELECTRIC DIGITAL DATA PROCESSING). Based on the patent features, we obtained the citation recommendations as shown in Table 6. In the output of Table 6, [O] denotes a correctly recommended patent, while [X] represents an incorrect one. Among the top 10 recommendations, seven patents were correct in the proposed model while only two patents were correct in the compared model. Additional examples of such qualitative results are presented in Appendix A. Although CRNet exhibits satisfactory prediction performance regarding the patents to be cited, this behavior degrades when there exist many overlapping words with the query patent.

**Table 6.** Qualitative result for CRNet (Example: US8612668B2)

| | Title | Abstract | CPC |
|---|---|---|---|
| Input | Storage optimization system based on object size | A method and apparatus optimizes storage on solid-state memory devices. The system aggregates object storage write requests. The system determines whether objects associated with the object storage requests that have been aggregated fit in a block of the solid-state memory device within a defined tolerance. Upon the aggregation of object storage write requests that fit in a block of the solid-state memory device, the system writes the objects associated with the aggregated object storage write requests to the solid-state memory device | G06F |
| Output (the proposed model) | 1. Storage Optimization System [X]<br>2. Method and apparatus for storing data in ash memory [O]<br>3. Method for ash memory data management [O]<br>4. Adaptive network content delivery system [O]<br>5. Multi-Operation Write Aggregator... [O]<br>6. Object-oriented cell directory database for a distributed computing environment [X]<br>7. Apparatus and methods for providing data synchronization... [X]<br>8. System and method for optimized access to memory devices requiring block writing [O]<br>9 Data Tree Storage Methods, Systems and Computer Program Products... [O]<br>10. Reliability of write operations to a non-volatile memory [O] | | |
| Output (the compared model) | 1. Storage Optimization System [X]<br>2. Method and system for network storage device failure protection and recovery [X]<br>3. Technique for increasing endurance of integrated circuit memory [O]<br>4. Storage system and data storage method using storage system [X]<br>5. Data write device and data write method [X]<br>6. Virtual incremental storage apparatus method and system [X]<br>7. Data storage system for providing redundant copies of data on different disk drives<br>8. Determining when to apply writes received to data units being transferred to a | | |

secondary storage [X]
9. Method for flash memory data management [O]
10. Managing writes received to data units that are being transferred to a secondary storage as part of a mirror relationship [X]

## 5. Conclusion and future works

In this paper, we addressed practically challenging issues for the patent citation recommendation task, which also includes lack of a high-quality public dataset. To this end, we proposed the PatentNet that was collected by patent examiners. Moreover, we also proposed a DNN-based patent citation recommendation method that can outperform other existing approaches using the PatentNet and strong benchmark model. However, this study suffers from a few drawbacks. First, the citation recommendation performance can be improved further. We employed a bag-of-words in our study, however, representation of context or relationship between words in a document is challenging. Second, the PatentNet contained a specific target field (such as geostationary orbit complex satellite), mainly due to the insufficient patent examples that were determined by examiners. Further validation regarding various fields is necessary for the adoption of the proposed method. Expanding the target fields of the proposed framework beyond the geostationary orbit complex satellite field also remains as another future research topic.

## Appendix
### A. Qualitative Results

**Table A1.** Qualitative result for CRNet (Example: US8475579B2)

|  | Title | Abstract | CPC |
|---|---|---|---|
| Input | Black ink composition | A black ink composition for post print and preprint, obtainable by adding (colour concentrate mixture) solid red concentrate, blue concentrate, and green concentrate to a carbon black ink having a concentration of 10-25% by weight of the composition, wherein the carbon block has not undergone oxidation treatment, and the colour pigments are made up 75-90% by weight. | C09D |
| Output | 1. Black ink composition [X]<br>2. Water-base ink for ink jet recording [O]<br>3. Cationic pigments and aqueous compositions containing same [O]<br>4. Ink composition [X]<br>5. Modified carbon products and ink jet inks, inks and coatings containing modified carbon products [O]<br>6. Reaction of carbon black with diazonium salts, resultant carbon black products and their uses [O]<br>7. Inks and other compositions incorporating limited quantities of solvent advantageously used in ink jetting applications [X]<br>8. Ink composition for ink-jet recording, recording method using same, and record [O]<br>9. Water-based pigment ink, and ink-jet recording method and instruments using the same [O]<br>10. Pigment inks having excellent image and storage properties [O] | | |

**Table A2.** Qualitative result for CRNet (Example: US8429002B2)

|  | Title | Abstract | CPC |
|---|---|---|---|
| Input | Approaches for enforcing best practices and policies | Approaches for employing computerized processes to conform the behavior of users to a set of best | G06Q |

|  | through computerized processes | practices and policies is provided. In an approach, a tool for use in performing a desired task may be identified. A plurality of characteristic values is stored, on a machine-readable medium, for each of a plurality of tools. Each of the plurality of characteristic values identifies a number that reflects how much the tool associated with the characteristic value supports a particular characteristic. A score for each of the plurality of tools is computed. The score for each tool is the sum of the characteristic values that are associated with each tool. Thereafter, display data that describes how to render a pictorial representation, which uses the scores in depicting a relative measure of how appropriate each of the plurality of tools is for the desired task, may be transmitted to a client. |  |
|---|---|---|---|
| Output | 1. Approaches for Enforcing Best Practices and Policies Through Computerized Processes [X] 2. Structured methodology and design patterns for web services [X] 3. Method for implementing a best practice idea [O] 4. Utilizing graphs to detect and resolve policy conflicts in a managed entity [O] 5. Method, system and tools for performing business-related planning [O] 6. Resource management planning [X] 7. Apparatus, system and method for automatically making operational selling decisions [O] 8. Business process outsourcing [O] 9. Testing practices assessment toolkit [O] 10. Comparatively analyzing vendors of components required for a web-based architecture [O] | | |

Acknowledgments